# Degradation Kinetics of Inverted Perovskite Solar Cells

Mejd Alsari[1*], Andrew J. Pearson[1], Jacob Tse-Wei Wang[2,3], Zhiping Wang[2], Augusto Montisci[4], Neil C. Greenham[1], Henry J. Snaith[2], Samuele Lilliu[5,6], Richard H. Friend[1*]

[1] Cavendish Laboratory, University of Cambridge, CB30HE Cambridge, UK
[2] Clarendon Laboratory, Department of Physics, University of Oxford, OX1 3PU Oxford, UK
[3] CSIRO Energy, Mayfield West, NSW 2304, Australia
[4] University of Cagliari, Dept. of Electrical and Electronic Engineering, 09123 Cagliari, Italy
[5] Department of Physics and Astronomy, University of Sheffield, S3 7RH Sheffield, UK
[6] The UAE Centre for Crystallography, UAE

* Correspondence to: rhf10@cam.ac.uk, ma671@cam.ac.uk

## Abstract

We explore the degradation behaviour under continuous illumination and direct oxygen exposure of inverted unencapsulated formamidinium(FA)$_{0.83}$Cs$_{0.17}$Pb(I$_{0.8}$Br$_{0.2}$)$_3$, CH$_3$NH$_3$PbI$_3$, and CH$_3$NH$_3$PbI$_{3-x}$Cl$_x$ perovskite solar cells. We continuously test the devices *in-situ* and *in-operando* with current-voltage sweeps, transient photocurrent, and transient photovoltage measurements, and find that degradation in the CH$_3$NH$_3$PbI$_{3-x}$Cl$_x$ solar cells due to oxygen exposure occurs over shorter timescales than FA$_{0.83}$Cs$_{0.17}$Pb(I$_{0.8}$Br$_{0.2}$)$_3$ mixed-cation devices. We attribute these oxygen-induced losses in the power conversion efficiencies to the formation of electron traps within the perovskite photoactive layer. Our results highlight that the formamidinium-caesium mixed-cation perovskites are much less sensitive to oxygen-induced degradation than the methylammonium-based perovskite cells, and that further improvements in perovskite solar cell stability should focus on the mitigation of trap generation during ageing.

The relatively high power conversion efficiency (PCE)[1] of perovskite solar cells (PSCs) combined with their potential for low-cost production[2] and their outstanding opto-electronic properties such as band-gap tuneability,[3] long charge diffusion length[4], low recombination rates,[5] and photon recycling,[6, 7] would make these devices ready for the PV market, although long-term stability remains a concern.[8] PSCs degradation can take place in the light-absorbing perovskite layer and/or in any intermediate layers, which can degrade due to their intrinsic structural instability and/or due to external factors, such as oxygen, moisture, heat, electrical bias, and mechanical stress.[9] Research into the degradation mechanisms of PSCs has so far predominantly focussed on regular n-i-p architectures. Inverted p-i-n devices can potentially outclass the n-i-p stack, both in terms of efficiency and stability, provided that stable *n*-type materials can be identified.[8] In this work, we explore the degradation kinetics of unencapsulated inverted p-i-n PSCs employing the benchmark $CH_3NH_3PbI_3$ and $CH_3NH_3PbI_{3-x}Cl_x$, and a more thermally durable alternative $FA_{0.83}Cs_{0.17}Pb(I_{0.8}Br_{0.2})_3$ perovskites as the photoactive layers.[10, 11]

Recently, we investigated the degradation kinetics of unencapsulated regular $CH_3NH_3PbI_{3-x}Cl_x$ (MAPIC) PSCs under continuous illumination in dry $N_2$ (stabilization phase) and $N_2:O_2$ (stress phase) atmospheres.[12] Current-voltage (IV) sweeps, transient photocurrent (TPC) and transient photovoltage (TPV) measurements were continuously and sequentially acquired *in-situ* and *in-operando*. During the stress phase the PCE was exponentially lost over time due to the emergence of a space-charge within the device that impeded charge extraction and accelerated photo-oxidation of the perovskite layer.[12] Here, we use the same setup to age MAPIC, $CH_3NH_3PbI_3$ (MAPI) and $FA_{0.83}Cs_{0.17}Pb(I_{0.8}Br_{0.2})_3$ (mixed-cation) PSCs. The intrinsic stability of MAPI is poor due to the volatility of the methylammonium (MA) cation.[13] As MA sublimates, the perovskite converts into $PbI_2$-rich domains that lower the efficiency of charge generation and impede charge transport between perovskite grains, thus affecting the open-circuit voltage ($V_{oc}$) and the short-circuit current ($J_{sc}$).[9] To overcome these issues, more structurally stable perovskites have been obtained by replacing the MA cation with complex cation mixtures.[14-16] The caesium/formamidinium (Cs/FA) combination has been used to fabricate structurally stable and band-gap tuneable $FA_{0.83}Cs_{0.17}Pb(I_xBr_{1-x})_3$ regular PSCs with relatively high PCEs .[11, 17, 18] Here we use the mixed-cation devices to provide a point of comparison between PSCs with active layers of differing intrinsic stability.

In Figure 1 we show the evolution of the normalized figures-of-merit (FOM) extracted from reverse and forward IV sweeps (Figure S1-Figure S3) of the three inverted devices stressed under continuous simulated solar illumination (AM 1.5 G) in dry $N_2$ and in dry $N_2$ (99%): $O_2$ (1%) atmospheres. All devices discussed here have the architecture FTO/PEDOT:PSS/Poly-TPD/Perovskite/PCBM/BCP/Au (see SI for Materials and Methods). Such a device structure results in a negligible hysteresis (Figure S4-Figure S6) compared to analogous regular n-i-p devices,[12] due to the good charge extraction properties of PCBM, and presumably fewer defects responsible for charge recombination at the perovskite charge extraction layer interface.[19]

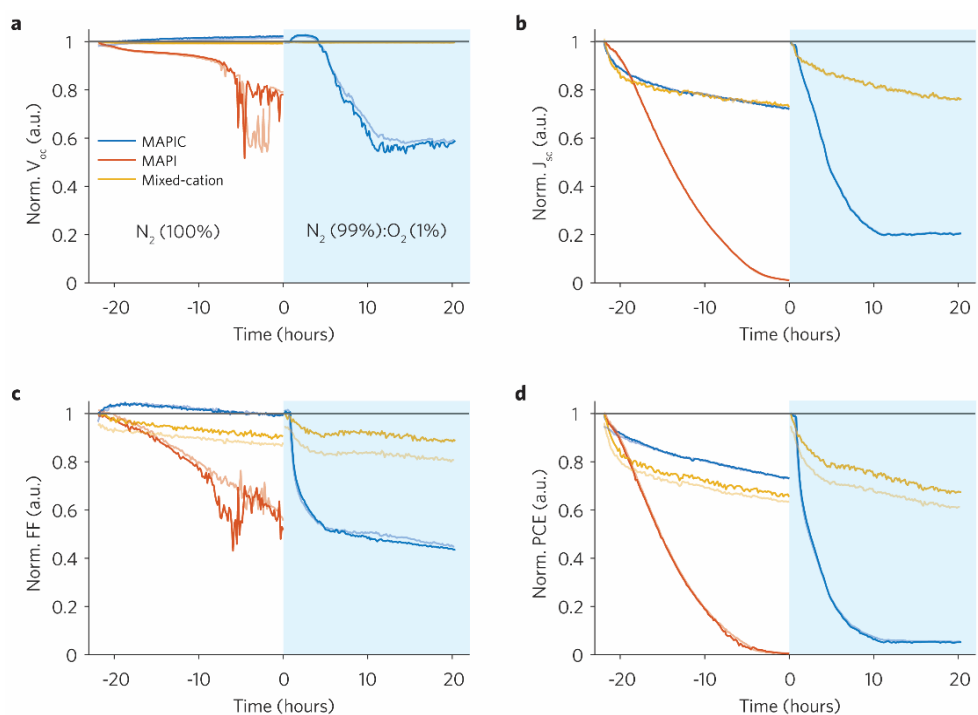

**Figure 1 | Evolution of the figures-of-merit (FOM) of inverted MAPIC, MAPI, and mixed-cation PSCs under continuous illumination and dry N$_2$ (Time < 0) and dry N$_2$ (99%) with O$_2$ (1%) (Time > 0).** Solid lines correspond to the FOM extracted from the IV reverse sweeps (from 1 V to 0 V). Pale lines represent forward IV sweeps (from 0 V to 1 V). Metrics during stabilization (Time < 0) and stress (Time > 0) phases are normalized to the first recorded value of the reverse metrics during stabilization and stress phases, respectively. Normalized open-circuit voltage $V_{oc}$ (**a**), short-circuit current $J_{sc}$ (**b**), fill factor FF (**c**) and power conversion efficiency PCE (**d**).

During the stabilization phase (Time < 0) all devices undergo a reduction in PCE, with the MAPI PSC experiencing total failure within 20 hours. The loss in the PCE of the MAPIC and mixed-cation PSCs is mainly due to a reduction in $J_{sc}$, however for the MAPI device, the $V_{oc}$ also reduces. The superior stability of MAPIC over MAPI during the stabilization phase could be an effect of PbCl$_2$ in the precursor solution resulting in a perovskite layer with improved morphology and/or lower defect density[20, 21], although the exact mechanism(s) for stability enhancement are not fully understood. During the stress phase (Time > 0) the MAPIC PSC completely degrades to ~5% of its initial PCE over 20 hours whilst the mixed-cation device retains ~70% of its initial PCE. The $V_{oc}$ of the mixed-cation PSC remains constant throughout both phases, suggesting that the perovskite remains stable and is not apparently affected by halide segregation. The MAPIC device turned into yellow colour at the end of the stress phase, consistent with the known mechanism for generation and reaction of superoxide (O$_2^-$), which subsequently decomposes the methylammonium halide within the perovskite crystal.[22] In general, for all the devices, most of the losses in the PCE are due to losses in the $J_{sc}$. Therefore, we postulate that photo-oxidization, or degradation of the charge extraction layers, or their interface with the perovskite could be playing a role with reducing the charge extraction efficiency.

To explore in detail the photocurrent loss mechanisms we consider the evolution in TPC and TPV traces measured in sequence with the IV scans during the stabilization and stress phases (see Figure S7-Figure S9 and Figure S13-Figure S18 for TPC and TPV data, respectively). From the TPC traces we identify five types of photocurrent behaviour, which are represented in Figure 2a. Slow components (dominant in type 3 and 4) are typically attributed to charge trapping/de-trapping and recombination processes, while fast transients (dominant in type 1, 2 and 5) are compatible with timescales associated with charge carrier transport.[23-25] The TPC dataset was clustered with a Pattern Recognition Neural Network (PRNN), which is a software-based computing system that works similarly to biological nervous systems. Once trained to recognize certain patterns, PRNNs can output fuzzy or intermediate answers. Here a PRNN (Figure S10) is trained with the TPC dataset shown in Figure 2a to provide a qualitative description of the TPC shape evolution during ageing. In Figure 2b we plot the extracted charge from the photocurrent decay transients and indicate the TPC curve types evolution during ageing. At the beginning of the stabilization phase all devices behave according to type 1 with a fast transient when the LED is switched on/off, which is indicative of the relatively clean and efficient photocurrent generation behaviour of the as-fabricated PSCs.[12] Continuous operation in $N_2$ induces changes in the TPC shape for all devices. The TPC of the MAPIC device transitions from type 1 to type 2 after ~6 hours of ageing. The photocurrent overshoot in type 2, observed in the first few μs of PSC illumination, may be attributed to the rapid formation of a transient diffusion gradient that enhances charge carrier recombination (reduces the photocurrent) before fading.[24, 26] The TPC of the MAPI PSC transitions from type 1 to type 4 after only ~3 hours of ageing, during which time the extracted charge from the photocurrent decays progressively reduces until the solar cell stops working. The TPC characteristics of the mixed-cation PSC immediately transitions from type 1 to type 5 after ~1 hour of ageing and maintains this behaviour until the last ~5 hours of the stress phase, when it goes back to type 1. Throughout ageing of the mixed-cation PSCs, the extracted charge experiences a negligible drop. In the MAPIC PSC, after ~1.5 hour exposure to oxygen the TPC transitions from type 2 to type 5 with a continuous decrease in the extracted charge. As the TPC traces further evolve from type 5 to type 3 the decay signal becomes negative, which is indicative of charge injection into the cell.[27] This observation and the photocurrent decay during the LED 'on' period could be explained by enhanced trap-assisted recombination and reduced charge de-trapping rate mechanisms.[28] While the charge density within the device increases due to continued photoexcitation, the competition between charge recombination and charge extraction in the PSC favours the former process to an extent that the steady-state photocurrent decreases. An increase in charge density within the PSC may also result in a space-charge that opposes the built-in field, resulting in a lower charge extraction efficiency.[12] For the mixed-cation PSC, the fact that the TPC shape does not seem to be influenced by the presence of oxygen indicates the superior stability of this device. We also observe that TPC type 4, seen during periods of severe photo-degradation, and type 3, which is dominant during the stress period, are both characterized by slow photocurrent decay transients (prolonged charge de-trapping and injection), compared to the other curves.

TPV measurements provide complementary information on the generation/recombination kinetics of photo-generated charges in the small perturbation regime.[27] In our degraded solar cells the TPV decays are best fitted with a double exponential function (see Figure S19-Figure S24).[20, 29] In Figure 2c we show that for the MAPIC device the fast time constant ($T_2$) dominates during stabilization ($a_2 > a_1$). During the stress phase however the slow time constant ($T_1$) increases and becomes dominant ($a_1 > a_2$) within ~5 hours before stabilizing. Although the origin of the slow and fast components are still under debate,[12, 29] we note that the double exponential behaviour is indicative of two populations of carriers that independently recombine.[20] In Figure 2d we show the evolution of the slow time constant $T_1$ versus $V_{oc}$ during the stress phase. This trend is compared to the ideal behaviour of the same device prior to ageing obtained by measuring TPVs under different light intensities. The $T_1$ vs $V_{oc}$ trend during the stress phase is non-linear with remarkably higher time constants compared to the ideal behaviour, suggesting that the time constants measured during the stress phase are likely to originate from trapped charges within the perovskite layer rather than free carriers. However, for the mixed-cation the dominant time constant is significantly lower (~1-2µs) and remains stable throughout ageing (Figure S22-Figure S24). This indicates that traps are not being generated in the perovskite layer and that the observed degradation might be due to degrading interlayers reducing current extraction and increasing the series resistance.

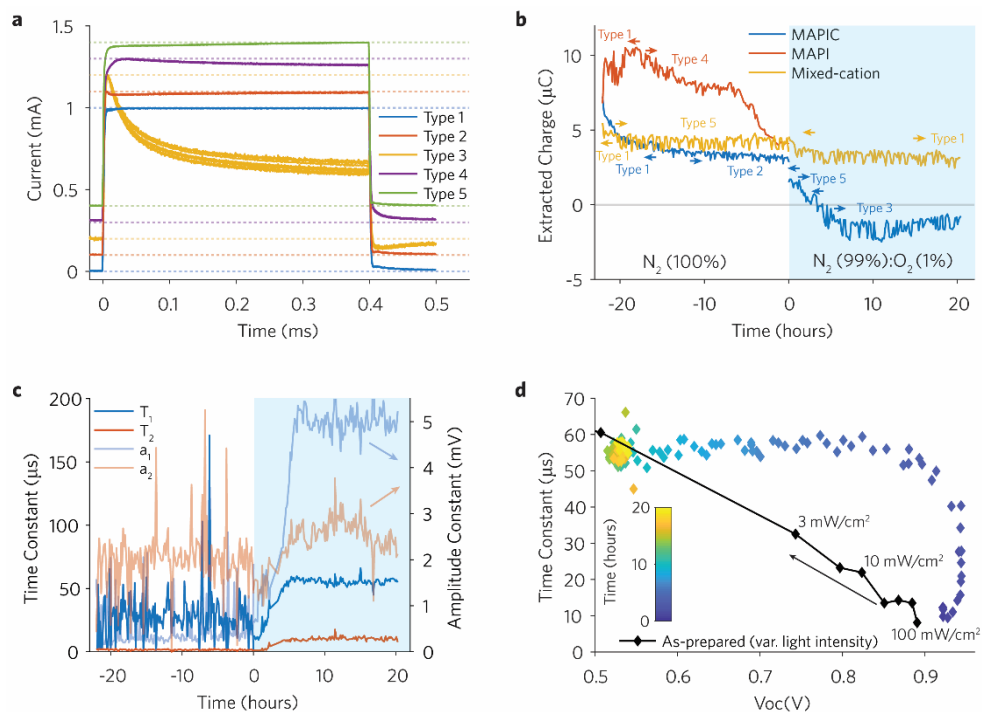

**Figure 2 | Stability kinetics extracted from transient photocurrent (TPC) and transient photovoltage (TPV) measurements of MAPIC, MAPI, and mixed-cation PSCs under continuous illumination and dry N₂ (Time < 0) and dry N₂ (99%) with O2 (1%) (Time > 0)**. **a**, Offset plot of types of behaviour for normalized TPC curves. **b**, Extracted charge obtained by integrating the TPC curves after the LED is switched off. The indicated TPC types are limited by arrows. **c**, Extracted time and amplitude constants from the double exponential fits of the TPV transient decays ($a_1 \times exp(-x/T_1) + a_2 \times exp(-x/T_2)$, where x is the time (µs)) for the MAPIC device. **d** Slow photovoltage decay time constant ($T_1$) versus $V_{oc}$ for the MAPIC device during the stress phase compared to reference values (black curve).

To further understand the recombination dynamics of the solar cells under stress conditions, we measured IV sweeps under variable light intensities (1-100 mW/cm² AM1.5G) before stabilization, at the end of stabilization, and at the end of the stress phase. The $V_{oc}$ versus the natural logarithm of the light intensity shows a linear behaviour (Figure S25) and from its slope $nkT/q$ we can extract the ideality factor $n$ (Figure 3).[30, 31] For the mixed-cation PSC $n \approx 1$ throughout ageing indicates bimolecular charge recombination.[32] For the MAPIC PSC the progressive increase of $n$ from ~1.66 to 2.53 during the stabilization phase indicates an increase in Shockley-Reed-Hall trap-based recombination. Further, we examine the power law dependence of $J_{sc}$ with light intensity ($J_{sc} \propto I^{\alpha}$) (Figure S26). The fitted alpha parameter (Figure 3) reduces throughout ageing for both MAPIC and the mixed-cation PSCs indicating the possible presence of trapped charges within the perovskite layer.[33]

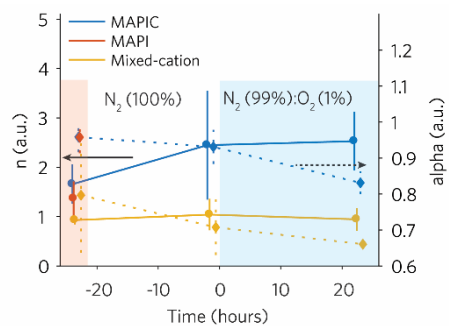

**Figure 3 | Ideality (n) and alpha factor extracted from $V_{oc}$ and $J_{sc}$ vs Light Intensity, respectively.** Measurements are performed before the beginning of the stabilization *phase*, at the end of the stabilization phase, and at the end of the stress phase. Full lines represent the extracted ideality factor, as indicated by the black arrow. Dotted lines represent the extracted alpha factor, as indicated by the black dotted arrow. Vertical lines indicate the 95% confidence intervals for *n* and alpha extracted from the fits.

In summary, we investigated the operational stability kinetics of unencapsulated $CH_3NH_3PbI_3$, $CH_3NH_3PbI_{3-x}Cl_x$ and $FA_{0.83}Cs_{0.17}Pb(I_{0.8}Br_{0.2})_3$ inverted perovskite solar cells in the presence of light and dry oxygen using *in-situ* and *in-operando* IV, TPC and TPV measurements. We confirm the superior stability of the mixed-cation PSCs compared to the benchmark PSCs. The observed light- and oxygen-induced degradation in the MAI-based solar cells occurs over shorter timescales than the mixed-cation devices, and is dominated by a loss in photocurrent and charge extraction efficiency. We interpret this to the generation of electron traps, resulting in long-lived trapped charge and the build-up of space-charge within the perovskite absorber layer. Our findings provide important insights towards understanding the operation of perovskite solar cells, and suggest that focussing on mitigating trap generation during ageing will lead to further improvements in perovskite solar cell operation.

## Acknowledgements

M.A. acknowledges the Ministry of Presidential Affairs (UAE) for supporting her doctoral studies. A.J.P. acknowledges support from the EPSRC through the grant EP/M024873/1. J.T.-W.W. acknowledges the EPSRC funding. We thank Chris Amey and Ravichandran Shivanna for preliminary measurements.

## Supporting Information

Supporting information is available at the following link:

https://www.dropbox.com/s/g3dvufiwzqyhskb/StressPerovskite_v4.1_SI.pdf?dl=0 .